\documentclass[openacc]{rstransa}

\begin{document}

\titlehead{Research}
\title{Gravitational lensing: towards combining the multi-messengers}

\author{
Anupreeta More$^{1,2}$,Hemantakumar Phurailatpam$^{3}$} 

\address{$^{1}$Inter-University Centre for Astronomy and Astrophysics, Ganeshkhind, Pune 411007, India.\\
$^{2}$Kavli Institute for the Physics and Mathematics of the Universe (WPI), University of Tokyo, 5-1-5,Kashiwa, Chiba 277-8583, Japan.\\
$^{3}$Department of Physics, The Chinese University of Hong Kong, Shatin, New Territories, Hong Kong\\
}

\subject{Cosmology, Observational Astronomy, Galaxies}


\corres{Anupreeta More\\
\email{anupreeta@iucaa.in}}

\begin{abstract}
The next generation of gravitational wave detectors and electromagnetic
telescopes are beckoning the onset of the multi-messenger era and the
exciting science that lies ahead. Multi-messenger strong gravitational
lensing will help probe some of the most important questions of the Universe in an
unprecedented manner. In particular, understanding the nature of
gravitational wave sources, the underlying physical processes and
mechanisms that produce emissions well before or right until the time of
the merger, their associations to the seemingly distinct populations of
gamma ray bursts, fast radio bursts and kilonovae. Not to mention,
multi-messenger lensing will offer unique probes of test of gravity
models and constraints on cosmological parameters complementary to other
probes.  Enabling multi-messenger science calls for concerted follow-up
efforts and development of new and shared resources required in the
community. 
\end{abstract}


\begin{fmtext}

\section{Introduction:}
We are currently in the era where two of the important predictions of
Einstein's theory of General relativity, Gravitational lensing and
Gravitational waves, have not only been observationally
confirmed but are being routinely discovered, studied and hold great
promise for the next few decades.
\end{fmtext}
\maketitle
Gravitational lensing is the phenomenon, due to the gravitational
potential of a massive object, which causes deflection of the light rays
coming from distant galaxies. In strong lensing, multiple, distorted,
(de-)magnified images of the distant background source can be seen under
suitable conditions. Since the light rays, 
corresponding to the multiple images, traverse different geometric path
lengths and experience different gravitational lens potential, the
observer sees a delay in the arrival times of these light rays. The
multiple images thus appear with some time delays between them. For general theory and review, see e.g.\cite{Schneider1992,Narayan1996,Meylan2006}.

Gravitational waves (GW) have enabled a novel way to study another dimension of the Universe.
Over the last decade, the GW detectors have discovered numerous GW events arising from compact binary coalescence systems, namely, binary black holes (BBH), binary neutron stars (BNS) or neutron star - black holes (NSBH) e.g.\cite{abbott2023gwtc,advancedLIGO2015,virgo2015,kagra2021}. Lensing of Gravitational waves was hypothesised and the
corresponding theory presented long before even the confirmation of
Gravitational waves e.g. \cite{PhysRevLett.77.2875,Takashi2003,lisa2011lensing}. The last few years has seen dedicated development and implementation of search algorithms to find lensed GW signals in data taken from the LIGO--Virgo--Kagra collaboration e.g.\cite{Hannuksela2019,LVlens2021,Liu2021,LVKlens2023,Li2023,McIsaac2020,dai2020searchlensedgravitationalwaves,Janquart2023}.

The detection of GW170817 is a remarkable moment in astrophysics, demonstrating the impact of multi-messenger astronomy in its truest sense
\cite{gw170817, Abbott2017}. It is the first observation of gravitational wave from a merging BNS. Besides, the signal was rapidly followed up by associated electromagnetic (EM) signatures across a wide spectrum, including gamma-ray bursts (GRBs), X-rays, ultraviolet, optical, infrared and radio waves e.g.\cite{LIGOScientific:2017ync,Alexander2017,Andreoni2017,Hallinan2017,Margutti2017}. 
Such multifaceted astrophysical information provided insights into the physics of neutron stars e.g.\cite{Kasliwal2017}, the speed of gravitational waves e.g. \cite{Liu2020}, and presented a new technique for measuring the expansion rate of the universe e.g. \cite{Abbott2017H0}.

This article attempts to summarise and highlight the importance of lensing in the multi-messenger studies that are either triggered by GW discoveries or help in better understanding sources that will produce GW. 

\section{Multi-messenger science with lensing}
There are numerous applications of combining the multi-messenger
observations which would be impossible otherwise. In this section, we
briefly describe a few such applications that have been explored and
stated in the literature. 

\subsection{Cosmology}
The standard cosmological model fits plethora of observations very well
although there are still a few outstanding problems to which there are
no clear explanations which leaves scope for alternate models to be
viable. For instance, the nature of dark matter. Lensing due to low-mass sub-galactic population of objects referred to as microlensing (lensing deflections$\sim\mathcal{O}(10^{-6})$~arcsec) and millilensing (lensing deflections$\sim\mathcal{O}(10^{-3})$~arcsec) arising from stars, stellar remnants, primordial black holes and dark matter clumps (or sub-halos) are ideal as the different cosmological models tend to make more distinct predictions for the properties of dark matter at smaller masses and smaller spatial scales e.g.\cite{More2009,Niikura2019,Diego2019,Basak2022,Vegetti2023}.

Another interesting problem is the tension in the measurement of the
Hubble constant ($H_0$) arising from differences in the constraints from the early Universe probes, for instance, cosmic microwave background measured by the Planck satellite ($H_0 = 67.4 \pm 0.5\ \mathrm{km/s/Mpc}$, \cite{Planck2020}) versus the late
Universe probes such as the distance ladder calibrated using Cepheids ($H_0 = 73.0 \pm 1.0\ \mathrm{km/s/Mpc}$, \cite{Riess2022}). 
\cite{Refsdal1964} proposed the idea that the measurement of time delays
between the multiply lensed images of a distant supernova could help in
constraining the Hubble Constant.
Historically, time-delay measurements of lensed quasars have emerged as a powerful cosmological probe. Studies, for example, from the TDCOSMO collaboration \cite{Wong2019,Birrer2021,Shajib2023,Wong2024} report an $H_0$ determination with precision at the level of $\sim 2\%$, for example, $H_0 = 65^{+23}_{-14}\ \mathrm{km/s/Mpc}$. 
These measurements are consistent with early-Universe probes but amplify the tension with late-Universe values. 
Lensed supernovae (SNe) have proved intractable to be discovered and be used for cosmology, owing to their low rate
of occurrence coupled with lack of suitable telescope surveys and observations, until very
recently e.g., \cite{Kelly2023,Pascale2024} although this is expected to change with upcoming surveys like Rubin-LSST e.g., \cite{Quimby2014,Goldstein19}.

Since lensed GW events will have extremely high precision in the time
delay measurements compared to optically lensed quasars and lensed SNe,
\cite{Liao2018} noted that cosmological constraints from lensed GW combined with EM data will
be far superior. Similarly, lensed FRBs are also expected to be excellent probes for high precision H0 measurements \cite{Li2018a}. The time delays between images in lensed GW and FRB systems can be measured with more precision and accuracy, often reaching the millisecond range, compared to the days (or hours, at best) for lensed quasars and supernovae. Furthermore, certain systematics such as
the uncertainty in the determination of the relative Fermat potential 
\footnote{The Fermat potential allows us to determine the arrival times of light rays or signals from a distant source which are affected due to the presence of an intervening gravitational potential. It depends on the extra geometrical path traversed by the light rays or signals and the lensing potential experienced by them. }, $\delta\Delta\psi=(\psi_{i_{\rm true}}-\psi_{i_{\rm mod}})-(\psi_{j_{\rm true}}-\psi_{j_{\rm mod}}$), at the
location of a pair of lensed images ($i,j$) are proposed to be
smaller which could lead to sub-percent precision on $H_0$ e.g., \cite{Liao2018}.

\begin{figure}[]
\centering\includegraphics[scale=0.4]{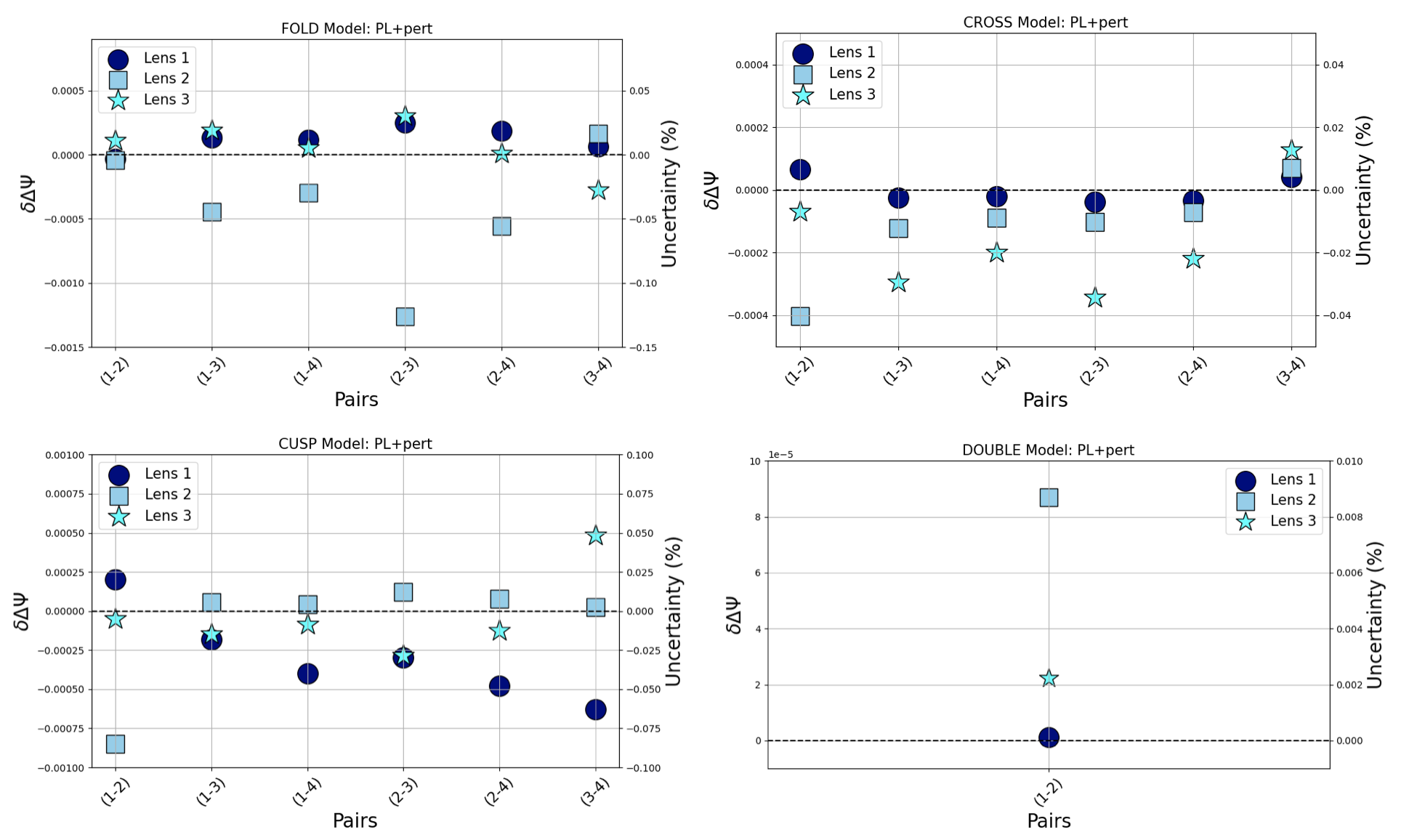}
\caption{The uncertainties in $\Delta\psi$ shown for different
combinations of pairs of lensed images for four lensed image
configurations (3-Quads: Fold, Cross and Cusp and 1-Double). In a fold and a cusp configuration, two and three of the four images, respectively, are nearly merging with each other whereas in a cross configuration, all of the four images are almost equidistant, similar to the ends of a cross symbol. Three lenses (Lens 1-3) are randomly generated to capture variations across different lenses per image configuration. The simulated lenses are assumed to follow a power-law mass density profile and a contribution from external shear or perturbation (PL+pert). The same model is assumed when finding the best-fit model although all of the parameters are kept free. In the subsequent iteration, this assumption will be relaxed which will lead to a further increase in the uncertainties (see Ali \& More 2025, in prep. for details). }
\label{fig:delpsi}
\end{figure}

Systematic investigations of actual uncertainties in the relative Fermat potential for
optical images, similar to the resolution of
the Hubble Space Telescope, are carried out (Ali \& More 2024, in prep.). 
Preliminary analyses indicate that the lens systems containing two images
(i.e. doubles) have smaller uncertainties compared those containing four
images (i.e. quads) as shown in Fig.~\ref{fig:delpsi}. Also, among quads, certain
configurations give much better accuracy in $\delta\Delta\psi$ across different combinations of image pairs. 
These results are tested and valid in idealised scenarios and are
consistent with the uncertainties stated in \cite{Liao2018} but are expected to increase substantially in realistic noise conditions. 

Strongly lensed point sources are almost always going to be affected by
microlensing by stars and stellar remnant population embedded in the
strong lens galaxy \cite{Dobler2006}.  Microlensing due to stellar and
stellar remnant population embedded in the strong lensing galaxy will
introduce modulations in the GW signals \cite{Mishra2021}. Both
parameter estimation of GW signals (e.g. the luminosity distance, chirp
masses and effective spins) and detection of strongly lensed GW signals
may get affected due to strong distortions produced by microlensing
\cite{Mishra2023}. Also, the fraction of strong lenses in which severe
microlensing (strong lensing magnification $>10$) are expected to be present are about 50\% for the current
GW detector sensitivities which will decrease for next generation
detectors \cite{Meena2022}. Thus, if the lens models fail to account
for the microlensing component, these may ultimately lead to certain
systematics in the inference of the cosmological parameters.

One interesting possibility to consider is that the microlensing
effects from the same stellar (-remnant) population will influence the
GW source as well as its EM counterpart, say the kilonova. However,
their imprints or observable effects will be different because, at GW
frequencies, the wave effects (interference and diffraction) will cause
frequency-dependent microlensing distortions whereas, at optical
frequencies, geometric-optics limit will apply and produce a constant
magnification due to microlensing for each of the strongly lensed image.
Therefore, joint analysis, combining the data from two messengers, will
give unique constraints on the microlens population. This can further
be combined with imaging of the lensed host galaxy (unaffected by
microlensing) to obtain accurate lens mass models which is essential for
doing precision cosmology.

\subsection{Test of General Relativity} 

Properties such as graviton mass, speed, polarisation and propagation of GW
are some of the tests of general relativity. Constraints on these
properties have been placed with the help of GW signals from compact binary coalescences. However, in presence of electromagnetic counterparts to GW sources and with lensing,
further interesting studies with improved constraints have been proposed,
as discussed below.

\subsubsection{Speed of Gravitational waves}
General relativity predicts that light and gravitational waves travel at
the same speed, take same path and experience the same Shapiro-time delay.

If a GW source with an EM counterpart is strongly lensed by a foreground
galaxy, then the time delays between multiple GW events can readily be
compared with the time delays between their respective multiple EM
images e.g. \cite{Collett2016,Fan2017}. By measuring the times of appearance of a pair of images from, say, EM observations and knowing the time of the first event in GW, the time of the second GW event can be predicted accurately. If the prediction does not match the observations then it may indicate a violation of GR.
Such a method will bypass the need of any knowledge of the
intrinsic time delays between the GW source and its EM counterpart. 

It is noted though that the arrival times of GW compared to their
EM counterparts may differ for lenses at $M<10^5M_\odot$ without violating general relativity \cite{Takahashi2017}. This happens because of
the longer wavelengths of GW signals compared to the deflections
produced by the low mass lenses wherein wave optics effects such as
diffraction and interference occur and the signal does not experience
a Shapiro time delay. For the EM counterpart, however,
one can use geometric optics approximation wherein the multiple images
are formed at the stationary points following Fermat's principle and the
images experience a Shapiro time delay. In such circumstances, there
will be a difference in the $\Delta t_{\rm GW}$ compared to $\Delta
t_{\rm EM}$ although when dealing with typical strong lensing effects by
foreground galaxies (masses much larger than $10^5M_\odot$), both the GW
and EM signals can be treated under geometric optics approximation and
any differences seen in the $\Delta t$ between GW and EM can be
attributed to departure from General relativity.

\subsubsection{Modified Gravitational wave propagation}

During the propagation of the GW signals over large cosmological
distances, their amplitudes may get attenuated by a different manner
unlike the expectation in GR for the amplitudes to be inversely proportional
to the scale factor. As a result, the luminosity distance measured via
the GW observations ($D_{\rm L,GW}$) of the source may not remain
consistent with its luminosity distance measured via electromagnetic
observations ($D_{\rm L,EM}$).  Thus, for alternate theories of gravity,
$D_{\rm L,GW}$ may show differences from $D_{\rm L,EM}$. For instance,
$D_{\rm L,GW} = D_{\rm L,EM} F(\theta)$ where the $F(\theta$) may
depend both on the parameters associated with deviation from GR models
and the cosmological parameters e.g.\cite{Finke2021,Narola2024}.

\begin{figure}[!h]
\centering\includegraphics[scale=0.4]{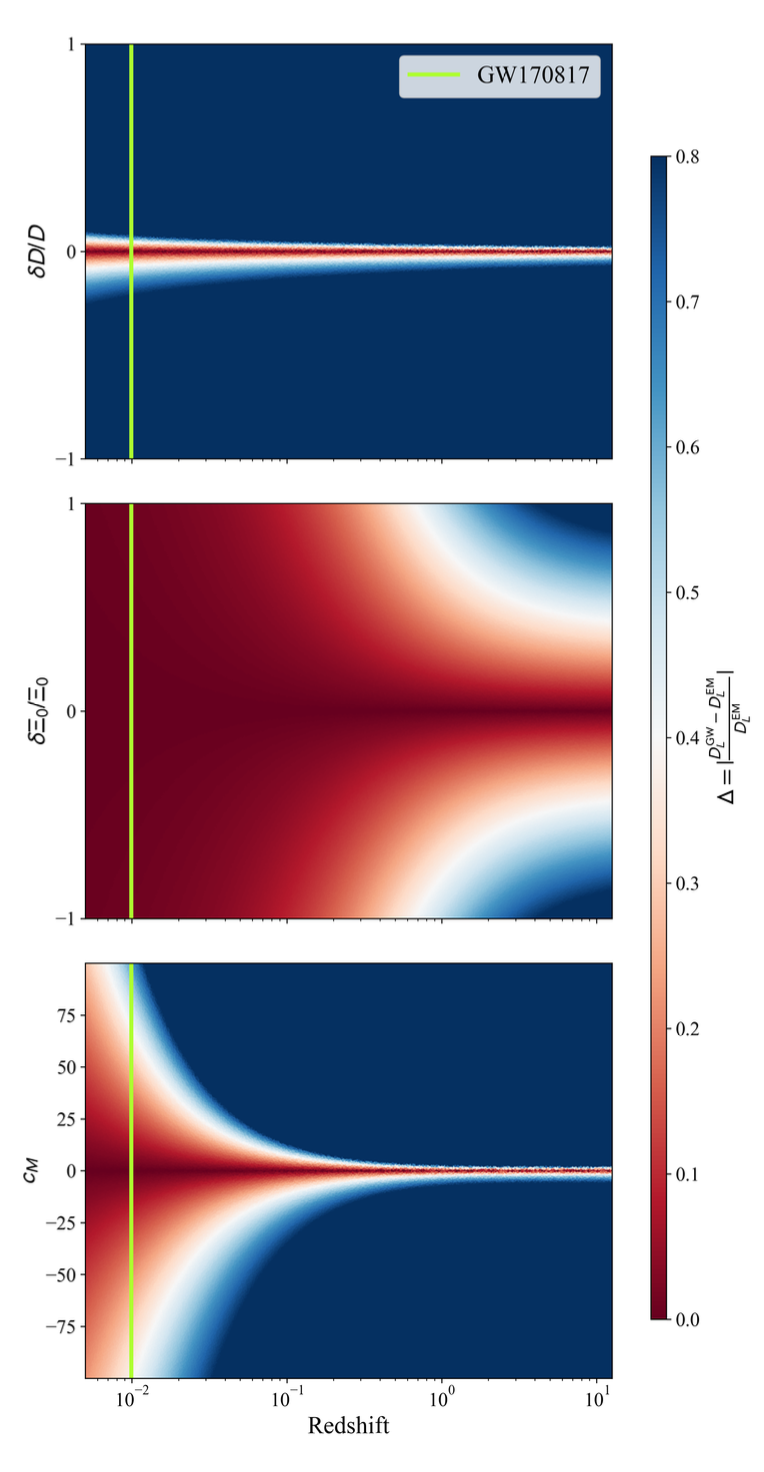}
\caption{The trends in $\Delta$ as a function of parameters of three
different non-GR gravity models (top, middle and bottom) and redshift.
The blue shaded regions indicate larger differences in $\Delta$ as
compared to red shaded regions and is dominantly seen at higher
redshifts for the parameters of most of the alternate models. Some model
parameters versus redshift show stronger contrast in the range of
$\Delta$ at all redshifts (top panel). Figure taken from \cite{Narola2024}. }
\label{fig:gw_prop}
\end{figure}

Suppose $\Delta$ denotes the relative difference in the luminosity
distances as $|D_{\rm L,GW}-D_{\rm L,EM}|/D_{\rm L,EM}$.
The dependence of $\Delta$ on the parameters corresponding to three
non-GR gravity models and on the redshift are shown in Fig.~\ref{fig:gw_prop} taken
from \cite{Narola2024}. These models are i) large extra spatial dimensions ($D$)
ii) a model that captures different propagation effects corresponding to
various theories of gravity alternative to general relativity ($\Xi$)
and iii) a running Planck mass ($c_M$). Further details of the
parametrization can be understood from \cite{Narola2024}. For reference, GW170817
is shown with a green vertical line. At such low redshifts, the
parameters need to deviate substantially to be able to discriminate
various models.

Whereas if the GW signals are strongly lensed then the lensing magnification
will enable detection of sources from much higher redshifts. 
For instance, the lensed BBH signals will typically originate from redshifts much higher than $z=1-2$. Such lensed signals might be better probes of the
modified GW propagation as they can probe cosmologically large
distances.  Since  the various models, in Fig.~\ref{fig:gw_prop}, show
larger differences in the predictions at higher redshifts, the lensed
BBH may have more discriminatory power in terms of constraining
different gravity models.  Given that BBHs are less likely to have
direct EM counterparts, it might prove observationally challenging to
use BBH for this proposed methodology which requires existence of detectable multi-messenger signals.

Contrarily, the detectable lensed BNSs, even though rarer than the lensed BBHs,
will likely have extremely high magnifications e.g. \cite{Smith2023,Magare2023}, making them also high-redshift detections compared to the unlensed BNS population.
Furthermore, the lensed BNSs are more likely to have detections in the
EM domain provided suitably deep, high angular resolution and prompt
follow-up observations are carried out e.g. \cite{Smith2023}. Given these arguments
lensed BNSs seem much more promising to pursue for the studies of modified GW
propagation.

\subsection{Understanding the nature and physical mechanisms of the
Sources}
\label{ssec:srcphys}

The EM sources such as the short Gamma Ray Bursts (sGRBs) \cite{Abbott2017GRB170817A}, Kilonovae
(KNe) \cite{Smartt2017} and Fast Radio Bursts (FRBs) (still speculative \cite{lorimer2018decadefastradiobursts}) are thought to be associated with
mergers of binary compact objects comprising one or both components as
neutron stars. If such events, or strongly lensed images of such events, are detected, then one can learn
about the i) nature of the progenitors responsible for the emission in
the multi-messenger domain ii) time delay in the emission between the
different messengers (e.g. GW vs GRBs/FRBs/Kilonovae) can give us
insights into the physical processes and their causal connections iii)
once the associated host galaxy is identified, detailed investigations
of the transient-host relation, their angular separations and global
properties of the host such as the star formation rate, metallicity and
stellar masses can be conducted.

\subsubsection{Gamma Ray Bursts}
GRBs are some of the most energetic and luminous objects observed in the
universe. In the prompt phase, gamma-rays are produced from the collimated
relativistic outflow that pushes through the interstellar medium and
is powered by a central engine. It is followed by an afterglow phase,
detectable right from the X-ray to radio wavelengths, arising from both
the expanding outflow as well as the continuous energy injection by the
central engine (see reviews e.g. \cite{DAvanzo2015,Meszaros2019}). 
Short GRBs (sGRBs), traditionally defined as bursts lasting $\le 2$ seconds, are associated with compact binary mergers (BNS or NSBH). This classification has evolved though, as events like GRB 211211A and GRB 230307A, lasting over 10 seconds, have been robustly linked to neutron star mergers (\cite{Rastinejad2022, Troja2022, Levan2023, Anderson2024, Yang2022, Gompertz2022}). However, for simplicity, this analysis focuses on sGRBs with durations $<2$ seconds.

Lensing of GRBs helps unveil the internal structure and mechanics of jets, as well as the composition and dynamics of ejected materials \cite{Takashi2003}. Gravitational lensing amplifies and replicates GRB light curves, uncovering subtle characteristics such as internal shocks, energy distributions, and jet dynamics that would otherwise remain hidden. It also offers a unique opportunity to test high-energy astrophysics theories \cite{Sironi2015}. Furthermore, since the lensed GRBs are detected from much higher redshifts, it gives us insights into the early universe's evolution, star formation rates, and rate of occurrence of GRBs e.g. \cite{Oguri2010}. \\

\subsubsection{Kilonovae}
KNe were long hypothesized to be the thermal transients associated
with the mergers of binaries visible in the optical and infrared wavelengths
e.g., \cite{Li1998,Rosswog1999,Metzger2015}.
In the ``r-process", the heating of the
non-relativistic outflows by the decays of heavy nuclei formed
through rapid neutron captures is detected as KNe e.g., 
\cite{Lattimer1974,Freiburghaus1999}. 
The KNe light curve is made up of emission from a bunch of ejecta masses
arising from dynamical ejecta (during the merger) or neutrino/magnetically
driven winds ejecta (remnant discs in the post-merger phase). The properties of the ejecta such as
their masses, velocities and opacities can be
determined using some of the binary parameters such as the chirp mass,
mass ratio, tidal deformability parameter and the equation-of-state
(EOS) e.g. \cite{Nicholl2021,Chunyang2023}.

Lensing will allow detection of KNe from higher redshifts $z>1$ whose
light curves will undergo cosmological time dilation making them
detectable for a longer period of time.  If early warning of the
upcoming lensed KNe images (and/or GW events) can be made, then one can
constrain the ejecta properties in the early evolutionary phases of KNe
which are impossible to without lensing until the era of third
generation GW detectors \cite{Magare2023}.  As a result, combined
analyses of lensed GWs and lensed KNe can help in better constraining
the EOS models of the component neutron stars and the degree of tidal
deformation found in the BNS/NSBH signals.

\subsubsection{Fast Radio Bursts}
FRBs are the milliseconds-duration radio signals, arising from
extra-galactic sources, dispersed by passage through an ionized plasma
in the line of sight. Even after the discovery of $\mathcal{O}$(100)
FRBs e.g. \cite{Lorimer2007,Thornton2013,CHIME2021}, their origins are still a mystery,
see reviews e.g. \cite{Cordes2019,Bailes2022}.  A small fraction
of the FRBs are found to be repeating with no particular periodicity
e.g. \cite{Kumar2019,CHIME2019} although it is
not confirmed yet that the remaining FRBs are genuinely not repeaters or
it is an observational bias. Some of the promising models for the
nature of the progenitors of the FRBs are i) a magnetized rotating young neutron star e.g. \cite{Bochenek2020,CHIME2020}) ii) BNS mergers - which may produce coherent radio emission by magnetic braking, similar to radio pulsars \cite{Totani2013}. Nevertheless, we note that the majority of FRBs cannot come from BNS mergers, since most FRBs seem to be repeating over much longer timescales than the lifetime of the merger remnant consistent with observations and reasonable theoretical models so far.

Quantifying the delay between the emission from FRB and GW sources (BNS
or NSBH) can help constrain the mechanisms responsible for the
production of FRBs which is much more feasible to achieve if the sources
are lensed \cite{Singh2023}. Similarly, different progenitor models lead to varied (lensed) FRB rates. Reversing the problem, the abundances of lensed FRBs could help identify which progenitor models are likely viable \cite{Li2014}. 
If there are multiply lensed images of a repeating FRB source with sufficiently high angular resolution (e.g. very large baseline interferometry in the radio), then any non-uniform motion inferred, based on accurate timing study, can reveal the nature of the FRB as well as the properties of its environments such as jet-medium interactions \cite{Dai2017}.

\subsection{Improved efficiency of searches}

Searches for the multi-messenger sources conducted with data from any one messenger alone may be less efficient than  searching for them jointly. It is suggested that if these sources are also lensed, there will be additional
constraints, leading to further improvement in the efficiency of their searches.
We recognise that lensing will be seen only for a small subset of multi-messenger sources but find it important to describe their advantages as given below.

\subsubsection{GWs-FRBs: Efficient discoveries with lensing}
As noted in Sec~\ref{ssec:srcphys}, the GW sources may potentially be
the progenitors of (subset of) the FRBs. Active searches to find
time-coincident events in these two domains are therefore needed
e.g. \cite{Wang2022,LIGOScientific:2022jpr}.
Searching for the association of FRBs with GWs is currently 
inefficient, primarily due to the following reasons.
Firstly, the delay time between the emission of an FRB from the time of GW event
is uncertain, especially, since not even a single confirmed association
exists.  
Thus, the time window within which counterparts should be searched are
unknown. Another factor is the lack of accurate distance measurements from both
GW e.g. \cite{Chassande2019}and FRB e.g. \cite{Xu2020} sources which introduces uncertainties in associating them
spatially in the direction of line-of-sight.
Lastly, the sky-localization uncertainties of the
GW sources, with current network of GW detectors, can vary from
$\mathcal{O}(10)-\mathcal{O}(100)$~sq.~deg. making it difficult to find
the true FRB counterpart even though FRBs have far better
angular resolution.

It is possible to overcome the above challenges if we are dealing with
lensed FRBs and GWs as suggested in \cite{Singh2023} owing to the
presence of repeated arrival of multiply lensed GW signals which will
have the same time delays in the associated lensed FRB images, provided
they are detectable. In some cases, even if one of the lensed
counterparts is weaker or fainter in one of the messengers (referred to
as subthreshold in GW), it may be possible to still discover the weaker
lensed counterpart based on the association in the data from the other
messenger (also, see Sec~\ref{ssec:grb_gw} below).

The top row of Fig.~\ref{fig:gw_frb} shows the time delay distribution 
for an intrinsic population of BNS mergers (blue curves) in comparison to
that for the detectable fraction from GW (orange curves) and the same
from FRB telescopes - CHIME (green, \cite{CHIME2018}) and BURSTT (red, \cite{lin2022}) telescopes. The
left panel shows the detectable population from current generation of
detectors e.g. the projected fifth observing run (O5) of
LIGO--Virgo--Kagra and the right panel from the future third generation
(3G) GW detectors and equivalent updated sensitivities to the FRB
telescopes. Not only the detected lensing fractions
increase with increasing sensitivity, the peak of the time delay
distribution also shifts to longer values due to higher redshift sources
entering the horizon \cite{Singh2023}. 

The false alarm probability (FAP) of randomly associating a detected
lensed FRB with the corresponding lensed GW pairs is found to scale
linearly with the number of discovered lensed FRBs \cite{Singh2023}. 
The FAP distributions arising from measurement
uncertainties in time delays are shown for increasing number of
discovered lensed FRBs (see bottom left panel of Fig.~\ref{fig:gw_frb}).
The FAP to obtain a correct association for 1 lensed FRB is about
$10^{-8}$. 

Even in the instances where the GW is being lensed by low mass lenses such
that the wave-optics effects will modulate the GW waveforms rather than
producing time-resolved multiple events, \cite{Singh2023} proposed that
the correct association between the GW and the FRB sources can be made.
An example of this is shown in the bottom right panel of
Fig.~\ref{fig:gw_frb} for the time delay posterior of a lensed NSBH
merger. The time measurement for FRBs is extremely accurate resulting in
a narrow posterior. Matching the independently measured time delays
between the lensed GWs and lensed FRBs can establish that they are
counterparts.

\begin{figure}[]
\centering\includegraphics[scale=0.6]{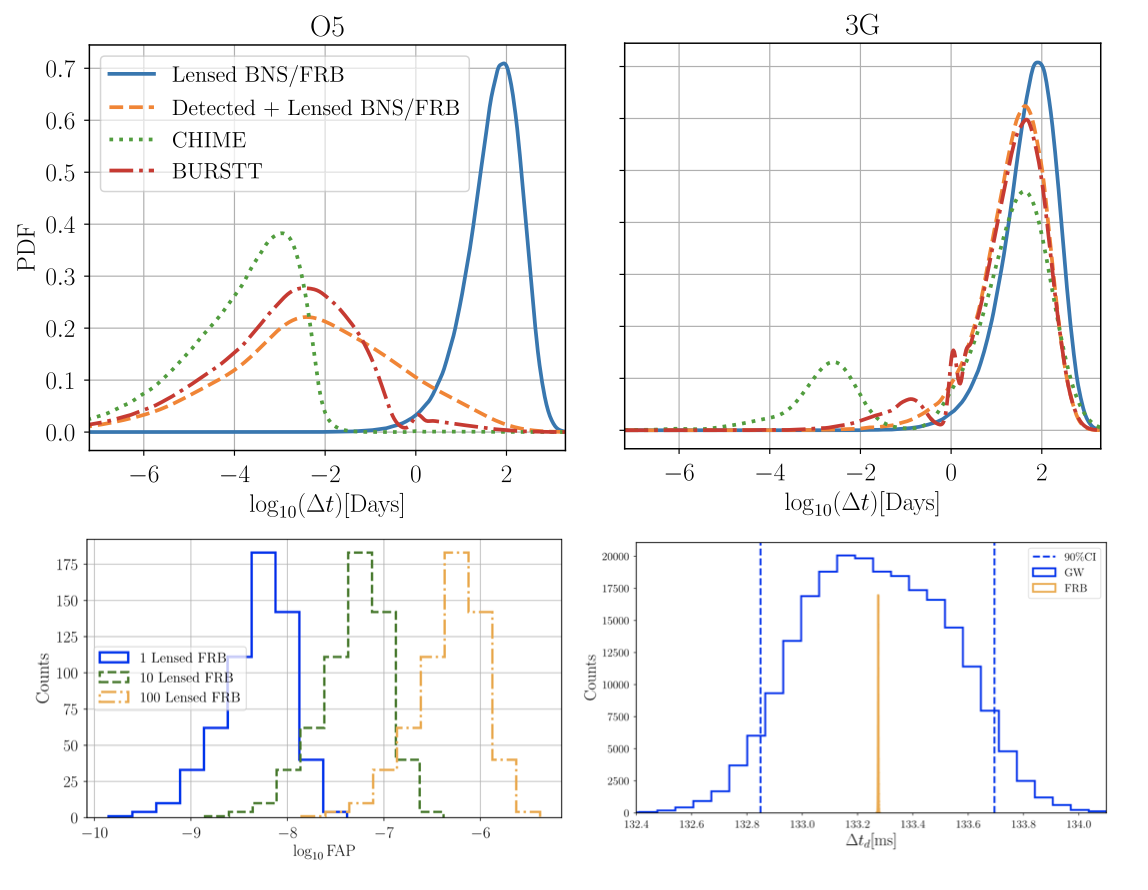}
\caption{
{\it Top:} Time delay distributions for lensed BNS mergers for
GW observing scenarios – O5 (left)  and
3G (right) including detectability from CHIME and BURSTT radio telescopes for
lensed FRBs. 
{\it Bottom Left:}Distribution of FAPs of lensed FRBs that can be associated with GW
signals arising from lensing GW time-delay uncertainties. 
{\it Bottom Right:} Time delay posterior inferred
for a microlensed NSBH (blue) and corresponding lensed FRB (orange). 
Figure taken from \cite{Singh2023}.}
\label{fig:gw_frb}
\end{figure}

\subsubsection{GRBs-GWs: Detecting sub-threshold GW counterparts}
\label{ssec:grb_gw}
Similar to the FRBs, the time-delay based association between the lensed
GRBs and lensed GWs can be extremely useful in detecting the
counterparts \footnote{In this section, GRBs and short
GRBs are used interchangeably to mean the same. Similarly, for GWs and
BNSs may be used interchangeably.}. 
There have been several efforts to identify lensed GRBs. Some studies involved analyses of decade-long data from the Fermi Gamma-ray Burst Monitor (GBM), examining around several hundreds of GRBs to identify potential lensed events \cite{Ahlgren2020,Chen2022}. Additionally, specific event(s) have been explored to find evidence of lensing in individual GRBs e.g. \cite{Wang2021,Yang2021}.

Two scenarios are presented for searches based on GRB-GW association 1) if lensed GRBs are detected then finding the
associated counterpart lensed GW events and 2) if lensed GW events are
detected then finding the associated lensed GRBs.  
It is assumed that all detectable sGRBs originate from BNS mergers and
that the luminosity produced is similar to that of GRB~170817A to
simplify the simulation. Additionally, the inclination angle of the GW
is expected to be aligned with the GRB jet axis
 \cite{Michael2024,Abbott2017}. This alignment is crucial as it
influences the detectability and characteristics of the observed GRB,
given the highly beamed nature of these emissions. 

For viewing angle $\theta<\theta_c$, i.e. within the core angle of
5~deg, all (unlensed) GRBs out to $z=5$ are expected to be detected by
Fermi and Swift ~\cite{Yuan2016}. For off-axis case, $\theta>\theta_c$,
the probability of detection becomes a function of $\theta$ and $z$ and
is given by 

\begin{equation}
P_{\rm det}(\theta,D_L) = \left\{ 
  \begin{array}{ c l }
    1 & \text{if } \theta \le 5 \text{ deg and } D_L \le 46652 \text{ Mpc } (z\sim 5) \\
    1 & \text{if } \frac{1.61\times 10^8}{4\pi D_L^2} \text{exp}(-\frac{\theta^{2}}{2\times 21.2^2}) \ge 1 \\
    0 & \text{otherwise.}
  \end{array}
\right.
\label{eq:pdet}
\end{equation}

The unlensed GRBs will only be detectable if $P_{\rm det}=1$ whereas for
the lensed GRBs, at least two of the images must satisfy the
detectability criteria.

Similarly, the (lensed or unlensed) GW population is considered
detectable for a network comprising the three LIGO--Virgo detectors,
operating at their ultimate design sensitivity, with a combined network
signal-to-noise ratio (SNR) greater than 6 (scenario 1) or greater than
8 (scenario 2). GW events with a network SNR greater than 8 are
considered super-threshold, while those with lower SNR are termed
sub-threshold events from the standard search pipelines. For lensed
populations, at least two events need to meet the detectability
condition.

\begin{figure}[]
\centering\includegraphics[scale=0.45]{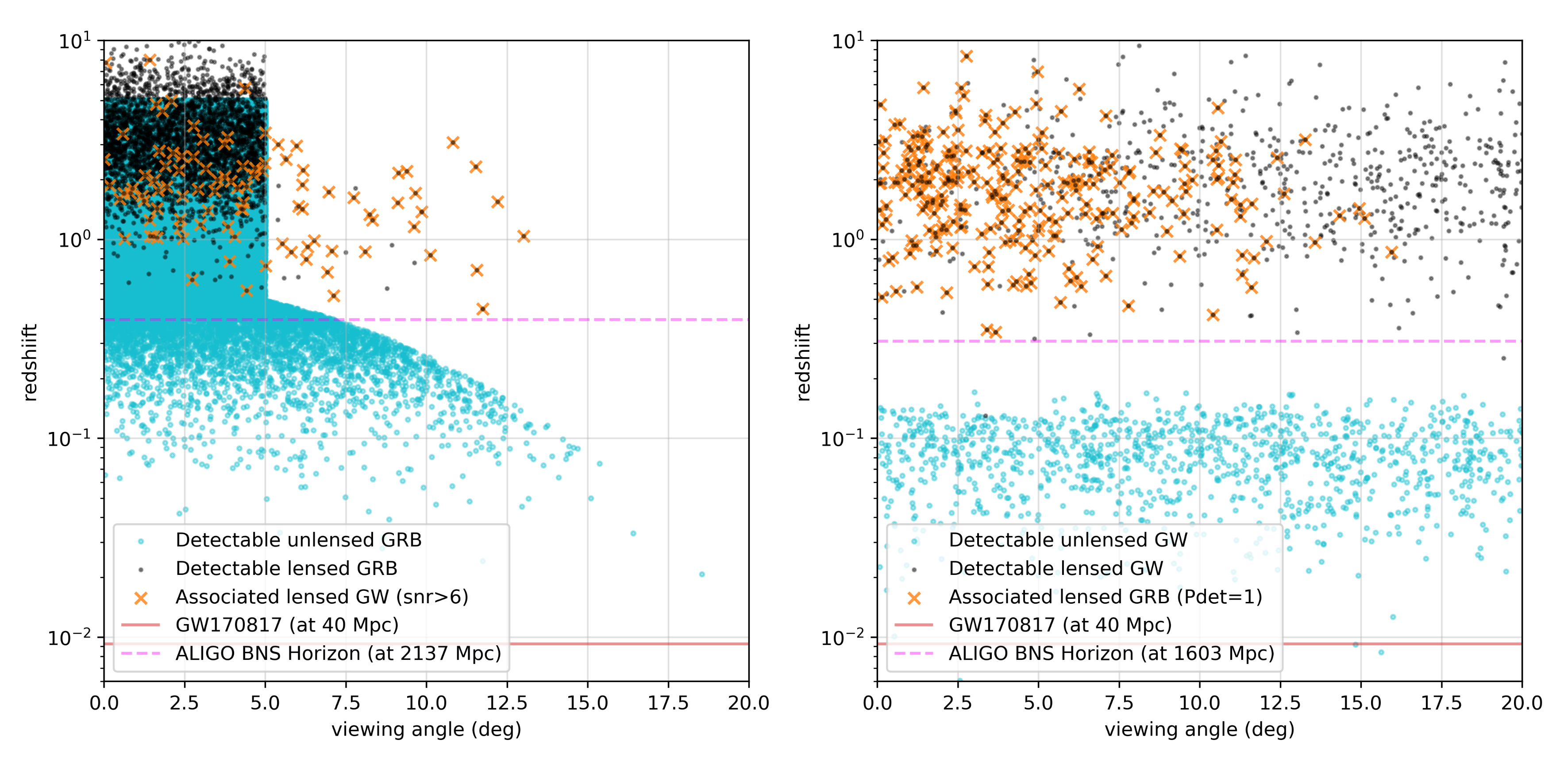}
\caption{Scenario 1 (left): Detectable lensed GRBs and the associated
detectable lensed GW events, lying in the sub/super-threshold
regime (SNR$>6$) at times, as a function of their redshifts and viewing
angle. For reference, the unlensed GRBs are also
shown along with the BNS horizon (assuming SNR$>6$) at Advanced LIGO design sensitivity
(purple dashed horizontal line) and the distance at which GW170817 was
located (red solid horizontal line). Scenario 2 (right): Similar to the
left panel although starting with detectable lensed GW events
(super-threshold with SNR$>8$) and their associated
detectable lensed GRBs. For reference, the unlensed GWs are also
shown along with the BNS horizon  (assuming SNR$>8$) at Advanced LIGO
design sensitivity (purple dashed horizontal line) and distance of GW170817 as
before (Phurailatpam et al.,in prep). See Sec.~2~d~\ref{ssec:grb_gw} for more details.
}
\label{fig:grb_gw}
\end{figure}

In Fig.~\ref{fig:grb_gw}, scenarios 1 (left) and 2 (right) show the
unlensed and lensed populations of GRBs and GWs along with the
associated counterparts in the redshift vs viewing angle parameter space
to give insights into their distributions. In scenario 1, the unlensed
GRB population is generated with Population I/II star merger-rate
density following \cite{Oguri2018,Wierda2021} and with the
luminosity function taken from \cite{Howell2019}.  The combined
detection with SWIFT and FERMI detectors is assumed to have a 50\% duty
cycle, meaning half of the sky is always visible. Through extensive
sampling and simulation, along with the application of the detectability
criteria, it is found that the number of GRB lens systems that meet the
detectability conditions is approximately 10~yr$^{-1}$. This assumes a
viewing angle range of $\theta_v\in[0,90]$ degree and redshifts
$z\in[0,10]$. Out of the detectable lens systems, 1 GW lens out of 1030
GRB lens systems in the sub/super-threshold domain (with SNR$>$6) could potentially be recovered with this approach.

In scenario 2, the detectable unlensed GW events are expected to be 21
for the Advanced LIGO design sensitivity over an observing run of
1~year. Applying the lensing rates \cite{ler2024, gwsnr2024}, about 1 GW lens
out of 400 unlensed GW events are detectable with the 3-detector network
SNR$>8$ (see right panel of Fig.~\ref{fig:grb_gw}).  Using the simulated
source and lens parameters for GW events, the corresponding $P_{\rm
det}$ value for the lensed GRBs is determined.  Searching for such
lensed GRB counterparts within a time window of 2~sec around the time of
the GW events, gives 1 GRB lens out of 13 GW lenses (see
\cite{hemanta2024grb} for detailed analysis).

\section{Towards enabling multi-messenger science}
\label{sec:enable_mma}
As some of the EM counterparts to the GW events occur within a short
time window or are sufficiently bright only for a limited timescales
(e.g. GRBs and KNe), concerted efforts among the multi-messenger
community and development of new tools, software and databases will be
imperative for enabling timely multi-messenger investigations and
analyses. Some ideas and thoughts in these directions are presented
below which are not exhaustive but useful to consider.

\subsection{Importance of additional GW detectors}

Multi-messenger studies depend critically on the angular resolution of
the GW observations. One of the biggest advantages of having additional
GW detectors in the network is the improvement in the sky localisation
\cite{Chu2012,Fairhurst2014, Saleem2022, Shukla2024} which is a decisive
factor in the efficacy of the successful EM follow-up observations. The
sky localisation uncertainties of the LIGO--Virgo network are
$\mathcal(O)10-\mathcal(O)1000$ sq.~deg.  The fraction of sources with
sky localisation of 20~sq.deg. increases by a factor of 1.6 (or 3) when
including a fourth GW detector in India (or a total of 5 GW detectors
with both LIGO-India and Kagra) \cite{Fairhurst2014}.

Furthermore, not only the sky coverage of the detected sample improves
i.e. the events can be detected from a larger fraction of the sky, at
any given time, but also the duty cycle of observations becomes better,
in other words, the fraction of time detectors are online in the network
within an observing run. More number of detectors also impacts the
sensitivity because one essentially obtains more observations of the
same event which can be combined (accounting for the antenna pattern) to
get better SNR, allowing signals from farther out to be detected
\cite{Fairhurst2014,Saleem2022}.  For example, comparison of two
networks i) two LIGO detectors (Hanford-H and L-Livingston) and ii)
three LIGO detectors (H, L and A-LIGO India) shows that the sky
localisation may improve by over 90\% for a
GW150914-like event, duty cycle by 20\% under multi-detector coincidence
SNR criteria and better sky coverage which results in a higher detection
rates of BBH signals by 40\%$-$50\% \cite{Saleem2022}.

Improved constraints on certain other GW parameters, for instance, the
luminosity distance and inclination angle becomes feasible
\cite{Saleem2022,Arun2014,Shukla2024} which are hampered otherwise owing to
inherent degeneracies. It is noted that many of the advantages of
having additional GW detectors are also achievable by a GW source that is lensed
since we get more observations of the same event at multiple epochs.
Regardless, in an expanded network of detectors, the lensed GW events are
not only benefitted by many of the aforementioned improvements but will also see 
reduction in the false positives owing to the improvements in sky localisation.
This can also have a positive impact on the search efficiency of the EM counterparts e.g., \cite{Yu2020,Wempe2024}. Moreover, the
improved constraints on the parameter estimation of the lensed GW
events will constrain the lens mass models better.

\subsection{Methodologies and tools for model predictions and analyses
in low latency}

Methodologies or tools that could help speed up and improve
efficiencies of follow-up studies in multi-messenger domains will either
need to be developed or are being developed. Some ideas are presented
below. 

\begin{figure}[]
\centering\includegraphics[scale=0.7]{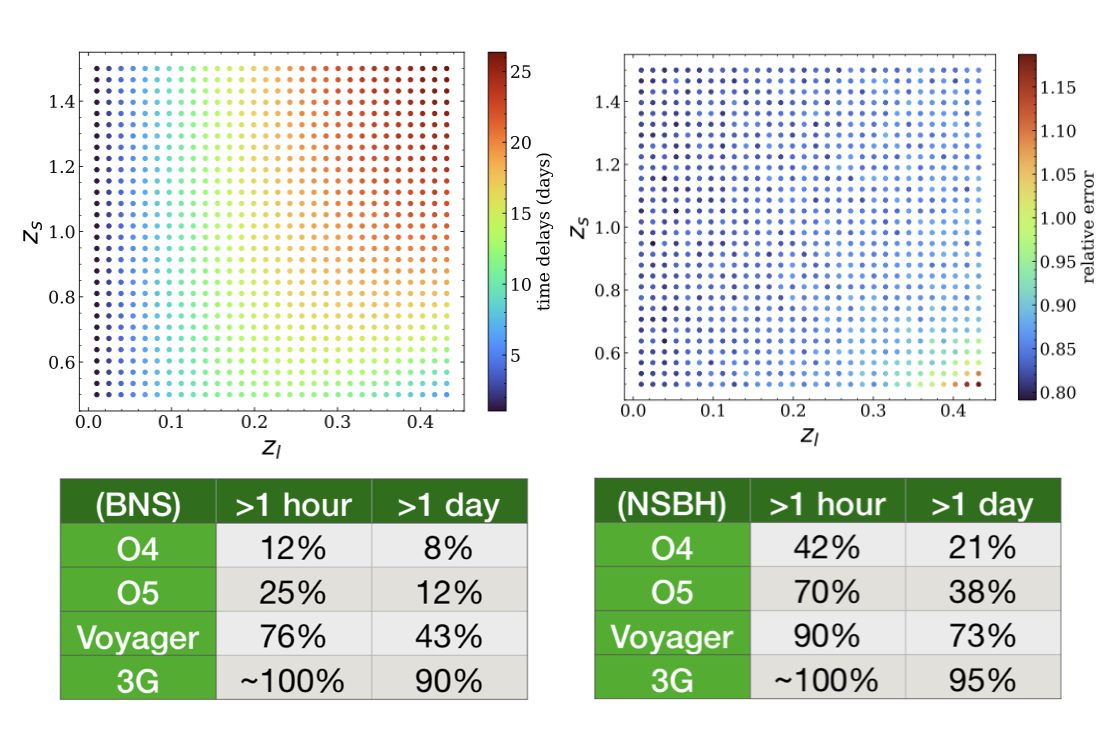}
\caption{ Time delay predictions and the relative errors as a function of the lens
and source redshifts (top row). The errors in the predicted time delays arise from the astrometric uncertainties in the optical lensed images. Fractions of lensed BNSs and lensed
NSBHs to have time delays above a threshold for various observing
scenarios (bottom row) such as the fourth and fifth observing runs of LIGO (O4 and O5), Voyager and 3G detectors (Einstein Telescope and Cosmic Explorer). Results are taken from \cite{Magare2023}.}
\label{fig:ew_lbns}
\end{figure}

While detection of GW170817 and subsequent detailed studies have
established BNS to be the progenitors of (some) sGRBs and KNe where the
latter are proved to be responsible for the production of the heavy
nuclei through the ``r-process'', the exact degree of contribution of
NSBH mergers to this universal r-process production is not yet well
understood which may vary from none to being the most dominant compared
to the BNSs. Only through timely multi-messenger studies, that provide
constraints on GW merger rates along with (non-)detection of the
counterpart EM sources, we will be able to ultimately place constraints
on the r-process yields. With NSBH as the progenitors, the KNe light
curves may not reach their peak at optical wavelengths for about a week
after the merger event, for some viewing angles. Therefore, early
observations of the rising light curves of KNe, particularly, when the
corresponding GW detection may not be informative, are of supreme
importance \cite{Gompertz2023}. Furthermore, early observations of the
emergent KNe, within about 2~days, are also critical in discriminating
between BNS and NSBH mergers in such cases. (Semi-)Analytical
frameworks have been developed for generation of KNe light curves based
on physical parameters of the BNS or NSBH signals in low latency
\cite{Nicholl2021,Gompertz2023}. As a result, rapid follow-up with
suitable observational set-up can be proposed for efficient detection of
the EM counterpart at optical and infrared wavelengths.

Similarly, as soon as a potentially lensed BNS or NSBH is detected in
the GW, it will be important to assess the plausibility of it being
lensed and rapid predictions of the properties of both the subsequent GW lensed
counterparts as well as the EM counterparts will be instrumental in the
design of the follow-up observations since the lensed source is likely
to arise from higher redshifts and the time delays between the subsequent
lensed events are more likely to be $\mathcal{O}$~hours-to-days
\cite{Magare2023} (see bottom row of Fig.~\ref{fig:ew_lbns}). Also, it is interesting to see that while the time delays are larger for lenses and sources situated at high redshifts, the relative time delay uncertainties do not change as dramatically as a function of these redshifts (top row of Fig.~\ref{fig:ew_lbns}).  Typical EM
follow-up imaging strategies which are intended for unlensed BNS/NSBH
may not go deep enough to detect the underlying host if it were actually
lensed. Moreover, without rapid and accurate lensing time delay
predictions, EM observations of the subsequent emergent KNe may not be
scheduled in the appropriate time windows. Thus, further studies are
ongoing that will help understand realistic uncertainties in the lensing
time delay predictions that could be obtained from the modelling of
ground-based (low angular resolution) imaging of any known (or mock)
lensed host galaxies. Additionally, methodologies and resources are
being developed which can actually process any newly discovered
candidate BNS/NSBH combined with early EM imaging (taken from the
archive or new EM data obtained with a latency of about 24~hours) to
make time delay predictions within a day or two .  

\subsection{Cross-matching investigations and Lens databases}
The GW (candidate) events triggered by the GW network of observatories
will have to be rapidly cross-matched with galaxy catalogs produced by
electromagnetic observations from existing surveys. The lens detection
pipelines will identify a pair of GW events to be promising candidates
for strong lensing.  The sky localisation contours of such pair of GW
events are expected to cover $\mathcal{O}(100)$~sq.~deg. in the sky.
There are many wide-area imaging surveys conducted in the EM, primarily,
in the optical, radio and infrared wavelengths with sufficient
sensitivity which would find hundred thousands of galaxies within the
sky localisation contours. Since many of these EM survey data also have
been searched for strong lensing, there are lensed EM galaxy (or
cluster) catalogs available in the literature (e.g. Master Lens Database
\cite{Moustakas2012}). 

Cross-matching the EM lens catalogs with the sky localisation regions
will produce $\mathcal{O}(100)$ matched lens systems.  Early studies
such as \cite{Yu2020,Hannuksela2020} have put forward some ideas on
localisation of the counterpart EM host galaxy, however, improved and
efficient methodologies are still needed.  Also, there are GW-centric
tools being developed such as
Lenscat\footnote{https://github.com/lenscat/lenscat}\cite{Vujeva2024}.  Further screening
of the matched lens systems can be done by extracting additional
observational constraints such as the redshifts, stellar velocity
dispersions or time delays. Alternatively, the imaging data can be used
for finding the lens mass models that best-fit matched lenses and their
corresponding predictions of time delays and image types (e.g. Type-I or
Type-II based on the topology of the Fermat's time delay surface \cite{Blandford1986}).  These data constraints will help in further vetting and
identification of the most likely candidates that are counterparts to
the pair of GW events e.g. \cite{Janquart2023}. A systematic large-scale modelling framework that can analyse
imaging data of heterogeneous angular resolution and depth for accurate
and generate data products including time delays, magnifications, image
types and other properties will be ideal.

\section{Summary and Conclusion}

Multi-messenger studies herald an exciting avenue of research that is expected to have an impact on areas of astrophysics, cosmology and fundamental physics. It will help unravel numerous mysteries such as the nature of progenitors of kilonovae and gamma ray bursts, contribution of neutron star - black holes in heavy nuclei
production,  potential association between gravitational wave sources
and fast radio bursts, alternate gravity theories, nature of dark
matter, Hubble tension, electromagnetic counterparts of supernovae and
almost certainly, completely unexpected discoveries in the next few
decades.

Strong lensing of gravitational waves or any of the electromagnetic
counterparts, in spite of their low occurrence rates, will push the
boundaries of our knowledge faster, for instance, by enabling certain investigations
sooner, by improving accuracy in certain models given their repeated
occurrences, by aiding in conducting optimised follow-up observations. Some of these ideas are touched upon in this article. A more
comprehensive description can be found in Smith et al. (this issue). In order to make this enterprise a successful one, efficient
coordination, cooperation and communication between the multi-messenger astronomy community will be essential. Moreover,
building resources, such as tools, pipelines and databases that are
publicly accessible, will maximise scientific output.

\vskip6pt

\ack{ The authors would like to thank Arshi Ali for sharing results from
an early version of the work.  We acknowledge generous support from The
Royal Society via funding of a Theo Murphy Meeting about Multi-messenger
Gravitational Lensing in Manchester, March 2024.  We also would like to
thank both the organisers of the meeting as well as the participants for
interactions and discussions which helped in preparation of this
article. Additionally, Hemantakumar Phurailatpam acknowledges support by grants from the Research Grants Council of Hong Kong (Project No. CUHK 14304622 and 14307923), the start-up grant from the Chinese University of Hong Kong, and the Direct Grant for Research from the Research Committee of The Chinese University of Hong Kong.
}


\bibliographystyle{RS}
\bibliography{mml_bib}
\end{document}